# Teaching Philosophy and Science of Space Exploration (PoSE)

Şerife Tekin
*University of Texas at San Antonio, One UTSA Circle, San Antonio, TX, 78249*
*University of Pittsburgh, Center for Philosophy of Science, 1117 Cathedral of Learning, 4200 Fifth Avenue, Pittsburgh, PA, 15260*

Carmen Fies
*University of Texas at San Antonio, One UTSA Circle, San Antonio, TX, 78249*

Chris Packham
*University of Texas at San Antonio, , One UTSA Circle, San Antonio, TX, 78249*
*National Astronomical Observatory of Japan, 2-21-1 Osawa, Mitaka, Tokyo 181-8588, Japan*

**Abstract**. Capitalizing on the enthusiasm about space science in the general public, our goal as an interdisciplinary group of scholars is to design and teach a new team-taught interdisciplinary course, "Philosophy and Science of Space Exploration (PoSE)" at the University of Texas at San Antonio (UTSA) where we currently teach. We believe that this course will not only help overcome disciplinary silos to advance our understanding of space and critically examine its ethical ramifications, but also will better educate the public on how science works and help overcome the science skepticism that has unfortunately become more prominent in recent years. In what follows, we first juxtapose two seemingly contradictory trends: increased interest in space science on the one hand and increased skepticism about and distrust in science on the other. We then turn to how our anticipated Philosophy and Science of Space Exploration (PoSE) course will develop tools that could dismantle distrust in science while also enhancing the scientific and philosophical understandings of space science. We explain the content and the questions we will examine in POSE and conclude with how we will measure our success and progress.

1. **Introduction**

Throughout human history, in certain periods, there has been a true inflection point in advancement. Consider the industrial revolution, the (first) space race, or the dawn of the information age. We are entering another such period – the rapid exploration of both nearby and outer space. NASA is planning a return to the Moon (~2024) and then Mars and very recently launched the $10B James Webb Space Telescope (December 25, 2021). New telescopes or probes could find evidence of life on distant exoplanets (planets around other stars) or solar system bodies (i.e., Mars, Europa, etc.) in the coming decades. SpaceX has already revolutionized the launch industry, with the global industry expected to be a $1.1 trillion market by 2040 (or much sooner). Mining asteroids, harvesting Moon resources, and militarizing space are planned. But what are the ethical implications of these activities? How can the problems of past "frontier" exploitations be avoided? How can we protect life, even microbial life, from terrestrial contamination (or annihilation), or should we care? What is the definition of life (there is no accepted definition), and what rights does it have? What are the ethics of billionaire joy-rides to space, given their climate footprint?

Capitalizing on the enthusiasm about space science in the general public, our goal as an interdisciplinary group of scholars is to design and teach a new team-taught interdisciplinary course, "Philosophy and Science of Space Exploration (PoSE)" at the University of Texas at San Antonio (UTSA) where we currently teach.[i] We believe that this course will not only help overcome disciplinary silos to advance our understanding of space and critically examine its ethical ramifications, but also will better educate the public on how science works and help overcome the science skepticism that has unfortunately become more prominent in recent years. In what follows, we first juxtapose two seemingly contradictory trends: increased interest in space science on the one hand and increased skepticism about and distrust in science on the other. We then turn to how our anticipated Philosophy and Science of Space Exploration (PoSE) course will develop tools that could dismantle distrust in science while also enhancing the scientific and philosophical understandings of space science. We explain the content and the questions we will examine in POSE and conclude with how we will measure our success and progress.

## 2. Distrust in Science vs. Increased Enthusiasm about Astronomy and Space Exploration

We are seeing two opposing trends in contemporary society with respect to trust in science. On the one hand, there is growing distrust in science in the United States. Even in the midst of the Covid-19 pandemic, the findings of medical experts on the spread of the virus and the effectiveness of vaccines have been dismissed or are met with skepticism by many around the world. Furthermore, there are many who resist the scientific consensus on climate change. Doubts about science and scientific findings have allegedly been sown by tobacco companies, free think-tanks, and other organizations whose economic interests and ideological commitments do not align with scientific findings (Oreskes, 2019; Oreskes and Conway 2010; Kitcher 2010). Trust in science is further shaken when scientific findings clash with world-views, religious commitments, or politics.

On the other hand, there is growing enthusiasm for and excitement about astronomy and space exploration (Fabian, 2010). Space science and astronomy are rigorous science areas, but in contrast to other disciplines, such as medicine or climate science, the public is incredibly enthused about these forms of inquiry. We are experiencing a rapid expansion in exploration of both nearby and outer space, and people are closely following these developments. For example, the launch of the James Webb Space Telescope generated much public interest, as evident in the many social media postings and videos, online launch watch parties, etc. There is much more to come, in terms of both space exploration and support from the public. Increased distrust in the areas of medicine and climate science is matched with increased trust in space exploration.

Our primary goal in PoSE is to harvest space science enthusiasm and use it to generate an educated and critical engagement with progress in space science, while at the same time addressing and challenging growing distrust in science. We can only overcome distrust in science through education and community engagement, and we believe POSE will help us achieve these goals. Finally, by reaching across typically separate colleges and departments in a university, we can broaden our reach to students who might not otherwise have the opportunity to study and consider such topics. In short, we hope our approach can broaden teaching of critical thinking to a wider community.

3. **Course Content and Teaching Methods**

To illustrate this contention, we will turn to history and philosophy of astronomy in PoSE, and consider what makes a theory scientific and distinguishes it from pseudoscience. Also known as demarcation criteria, many philosophers have examined the conditions under which a theory is ranked as scientific, or the criteria for deeming a particular theory scientific. In these discussions, astronomy almost always takes center stage as offering examples.

For example, Karl Popper, in an attempt to distinguish science from pseudoscience, argues that what makes a theory scientific is its "falsifiability" (Popper, 1959). This means every genuine *test* of a theory is an attempt to falsify it or to refute it. Testability is falsifiability; but there are degrees of testability. Some theories are more testable, more exposed to refutation, than others; they take, as it were, greater risks. An important example to explain what Popper means is the geocentric model of the universe, according to which Earth is assumed to be at the center of the solar system. One of the most developed geocentric models was that proposed by Ptolemy of Alexandria ($2^{nd}$ Century CE), an Egyptian philosopher, astronomer, and mathematician. Based on the observations he made with plain eye, Ptolemy perceived the Universe as a collection of nested and transparent spheres with Earth in the center. He postulated that the Moon, Mercury, Venus, and the Sun all revolve around Earth. Ptolemy's geocentric model of the universe is a genuinely scientific theory, from a Popperian perspective, because it is testable and falsifiable. In fact, scientists have successfully falsified the geocentric view, replacing it with the heliocentric model. More specifically, Nicolas Copernicus found that the geocentric model could not explain observational data while a heliocentric model could. The Sun – not Earth – was the center of the solar system. Galileo came to the same conclusion when, through the use of his telescope, he observed that Venus went through phases similar to Earth's moon. The nature of these phases was such that they could only be explained by Venus going around the Sun, not the Earth. What makes the geocentric model scientific, albeit false, according to the Popperian view, is its falsifiability with more empirical evidence and observation. We will further consider philosophical theories of induction in science and the meditations of Descartes. The underpinnings of these and other philosophical works can help to improve the thinking of scientists through an exploration of the very basis of critical thinking and scientific reasoning.

PoSE will take such historical and philosophical conversations a step further by referring to the recent advances in astronomy. As the above example shows, human interest in discovering the sky or the heavens is as old as the history of humani-ty. Questions about space have been debated by scientists, philosophers, and novelists for thousands of years, but only now do we have the science and advanced technology to address them. Space travel is no longer just an element of science fiction or a futuristic fantasy. Over the course of the last 60 years, since the first cosmonaut was launched into space in 1961, there have been tremendous scientific and technological advancements in space exploration, some of which are mentioned above. The feasibility of space exploration even by regular citizens will help humans answer some long-standing questions about humans' place in the Universe, the existence of (other) forms of life in space, and the history of our solar system.

However, space exploration also raises important questions that go beyond the bounds of scientific reasoning, requiring us to think critically about important epistemological and ethical questions. Consider a few additional questions to those mentioned in the introduction: "Who 'owns' the Moon? Or the planets?"; "Do humans have the right to 'conquer' other planets?"; "What are the impacts of colonization of a moon or of another planet"; "Why invest in space?"; "Could planetary exploration activities pose a contamination risk; if so, what kind of precautions must we adopt?" It is important to highlight that the basis of our program is to use the inspirational nature of these questions. As Bertrand Russell, a philosopher of science, has pointed out, science cannot be a boon or benefit society if it "precipitates evils greater than any that humankind has ever experienced" (Russell, 1948). Thus, the course is committed to helping students understand the importance of ethical practices in scientific research, thereby preparing them to be ethical and responsible researchers and citizens of and in the future.

In teaching PoSE, we will address these and other questions by bringing the tools provided by philosophy of science and ethics in conversation with scientific research in astronomy and pedagogical frameworks offered by STEM education. We will marshal the conceptual and empirical frameworks developed by each of these disciplines to encourage students to be informed consumers of space exploration. The cross-pollination between education, philosophy, and astronomy is currently suboptimal, but our hope is that our course will be a good example of how to escape these "silos."

The course will: (i) develop a curriculum in philosophy of space exploration that promotes the principles of equity, accessibility, and inclusivity; (ii) educate students on the complex scientific, and ethical questions pertaining to space exploration; (iii) help students develop distinctive and critical perspectives to examine and respond to such questions through methods of student-centered and experimental learning; and (iv) design authentic assessments that spark creative thinking and foster creative problem solving, while encouraging students to become lifelong learners by teaching them how to evaluate their own and their peers' performance. For example, we see this as an opportunity for STEM students who will become teachers; the course provides a location where they can think critically and deeply about the impacts and challenges related to the changing landscape of the scientific endeavor.

The project has the potential to have a transformative impact on teaching and learning, especially given our (physical and/or collaborative) proximity to NASA's Johnson Space Center (3.5 hr. drive from UTSA), NASA's Marshall Space Flight Center, the Space Telescope Science Institute, and SpaceX's Boca Chica Moon/Mars port. This yields numerous advantages, creating different and innovative learning opportunities for our students in the form of possible visits to the space science center or to SpaceX sites. We believe our diverse student population will also attract NASA's recruitment efforts, given their support of education efforts in STEM with an emphasis on increasing diversity in NASA's future. A recent report, *Science and Engineering Indicators*, the National Science Foundation (NSF)'s biennial assessment of global research, has highlighted serious—and persistent—inequities in science education, which has led them to emphasize the need for NSF to continue supporting efforts to improve science teaching (National Science Board, National Science Foundation 2022). Thus, our hope is that PoSE will also aid NSF's efforts to increase diversity in science.

In addition, we are working with UTSA's Academic Innovation to create an exemplary course for UTSA students in terms of content and interdisciplinarity, but also in terms of student engagement and digital fluency. Content will be created using the most innovative digital tools, such as Adobe Creative Cloud Apps and PlayPosit, and students will be engaged in the creation of interactive presentations. A focus on teaching responsiveness and digital accessibility will be a priority for the course design and delivery.

## 4. Measurement of Success

We intend to use the following metrics and evidence to evaluate the course's success. First, we will focus on making sure students develop a full grasp of the content covered. The extent of knowledge gains will follow a quantitative approach with pre- and posttest results, as well as a qualitative assessment of content acquisition through artifact and interview data. Second, we will measure the development of critical thinking using qualitative measures. Here, analysis of students' journal entries and papers as well as their contributions in classroom presentations will focus on how students engage with and interpret new information individually and in collaboration. For this, we will use rubrics that track the development of students' skills over time. Third, we will measure the change in students' interest in the subject matter via pre and post course surveys and interviews. We plan to triangulate findings to identify the degree of coherence in the analytic set. Finally, we will develop measures of the students' understanding of the tenets of ethics and of astronomy. We plan to organize an in-class conference for students, where each student presents a paper and comments on another student's paper. The intent is to support them in learning how to critically and productively engage with each other's ideas even if they do not agree with them. For example, we may ask students to imagine themselves as a life form on another planet when humans arrive; from that perspective, what do they perceive as appropriate measures. In evaluating their assignments, we plan to use a rubric used in "Ethics Bowls" to determine the strength of their critical reasoning.

## 5. Conclusion

Our hope in PoSE is to ignite further interest in scientific and ethical issues surrounding astronomy and space exploration and to illustrate how an interdisciplinary ap-proach to teaching, in this case a collaboration among philosophers, astrophysicists, and STEM educators and a cross-fertilization of the strengths of each of these disciplines, has much to offer to science and science education.

---

[i] More information on the course can be found on the course website, https://spark.adobe.com/page/wLiY8zK0tU3Tu/.